\documentclass[epj,final]{svjour}

\usepackage{latexsym}
\usepackage{url}
\usepackage{amsfonts}

\RequirePackage{graphicx}

\begin{document}
\title{CCC and the Fermi paradox}
\author{V. G. Gurzadyan\inst{1,2} \and R. Penrose\inst{3}% etc
}                     % Do not remove
\institute{SIA, Sapienza University of Rome, Rome, Italy \and Center for Cosmology and Astrophysics, Alikhanian National Laboratory and Yerevan State University, Yerevan, Armenia \and 
Mathematical Institute, Oxford OX2 6GG, U.K.}
\date{Received: date / Revised version: date}

\abstract{
Within the scheme of conformal cyclic cosmology (CCC), information can be transmitted from aeon to aeon. Accordingly, the "Fermi paradox" and the SETI programme  - of communication by remote civilizations - may be examined from a novel perspective: such information could, in principle, be encoded in the cosmic microwave background. The current empirical status of CCC is also discussed.   
\PACS{
      {98.80.-k}{Cosmology}   %\and
      %{PACS-key}{discribing text of that key}
     } % end of PACS codes
} %end of abstract

\maketitle
\section*{Introduction}

The so-called "Fermi paradox" \cite{D} refers to a puzzle that arises from an expectation that our own civilization is unlikely to have been the first to have arisen throughout our galactic neighbourhood, and if ours were not the first then, owing to the randomness involved in the timing of factors that lead to its development, the likelihood would have been that our civilization would have been preceded by others having an advantage of thousands of our centuries of technological development. The expectation, then, would be that such enormously advanced civilizations would have had ample opportunity to have either visited us or, at least, sent decipherable signals to us by now. The SETI programme \cite{D} has, for many years, been set up to detect such signals, but with no success as yet. This seeming puzzle of silence ("Where are they?") continues to attract attention, and there is a considerable variety of viewpoints and different approaches aimed at resolving this issue \cite{Webb}, spanning from the argument of the local uniqueness of our civilization up to a number of sophisticated unobservability schemes. This reflects, on the one hand, the unusual breadth of the topics raised by this seeming paradox and, on the other hand, the essential uncertainties in our knowledge of the many parameters involved.

In this note, we draw attention to a completely different aspect of this problem, which is raised by the scheme of conformal cyclic cosmology (CCC)\cite{Penrose2010}, which provides a new view on the origin and evolution of the Universe. Although CCC was first put forward about a decade ago \cite{Penrose2006}, it is only comparatively recently that observational evidence has come to light \cite{Gurzadyan2013}\cite{Meissner}\cite{An}, which appears to support some of the theoretical implications of CCC, and which seems hard to accommodate within the standard  $\Lambda CDM$ inflationary model. According to CCC, what is currently regarded as the entire history of our Universe, from its Big Bang origin to its infinitely exponentially expanding ultimate future is but a single aeon in an unending succession of broadly similar such aeons. In CCC, the current aeon is very similar to the picture presented by the  $\Lambda CDM$ model, differing from it primarily in that the early inflationary phase of  $\Lambda CDM$ (assumed to have occurred in a period between 10$^{-36}$ and 10$^{-32}$ seconds following the Big Bang) is taken not to be a feature of our current aeon, but whose effects arose, instead, from the ultimate exponential expansion of the aeon prior to ours \cite{Penrose2010}\cite{Gurzadyan2013}. In CCC, there is no contracting phase, but the transition from the ultimate expansion of one aeon to the big bang of the next is regarded as having occurred via an intermediate phase in which the material contents of the Universe consists solely of what are, in effect, massless particles satisfying a conformally invariant dynamics. This leads to an absence of an effective scaling (of both temporal or spatial dimensions) in this transitional phase, though the retaining of causal structure (well-defined null cones) allows the indefinitely expanding remote future of the previous aeon to be joined conformally smoothly to the big bang of the succeeding one. Well-defined dynamical equations allow this transition to take place in a deterministic fashion (though there remain some relatively minor unresolved issues in this dynamical evolution \cite{Penrose2010}\cite{Gurzadyan2013}.

\section{Black-hole collision signals from the previous aeon} 

An important feature of CCC is that occurrences that took place in the aeon prior to ours can have observable consequences in our own aeon. In earlier work \cite{Penrose2010}\cite{Gurzadyan2013} (see also \cite{Meissner}) we concentrated on the most violent processes that would have presumably taken place in that earlier aeon, namely collisions between supermassive black holes. The gravitational-wave bursts from such events would have released vast amounts of energy in a relatively tiny period of time and, according to CCC, these would have resulted in circular features in our CMB sky around which the temperature variance should be distinctively lower than average \cite{Gurzadyan2013} and the mean temperature either distinctively warmer or distinctively lower than the average, depending upon whether the source is, respectively, extremely distant or relatively close by. The angular radius of such circular rings could be anything up to around 20$^{\circ}$, or so, in the CMB sky \cite{Penrose2010}\cite{Tod}\cite{Nelson}, but the ring width would be expected to be generally around 4$^{\circ}$, with a minimum of about 2$^{\circ}$. Moreover, since such encounters between supermassive black holes would be expected to be repeated events within a single galactic cluster, several such rings would often appear in concentric sets. See \cite{Gurzadyan2013} for details, where we also show that there is some definite evidence for the actual presence of such rings in the WMAP data. It is worth noting that in an independent study \cite{DA} has recently confirmed the presence of the essential features in the WMAP data originally pointed out in \cite{Gurzadyan2013}, but where the authors argue that such features could consistently arise within the scope of the conventional  $\Lambda CDM$. In \cite{Gurzadyan2013}, we also showed, however, that these features critically depend upon their circular shape (as predicted by CCC), and that if the same analysis is applied to searches for elliptical rings, rather than circular ones, then the numbers fall off dramatically with increasing ellipticity (a feature also noted in \cite{An}).

A particularly striking feature of this analysis was the appearance of strong departure from homogeneity, where the ring centres are seen to be concentrated markedly in certain regions of the CMB sky. It should be made clear that although this feature was not a {\it prediction} of CCC, it is very well accommodated within that scheme, since the aeon preceding ours could well have possessed certain regions where there were extremely large galactic superclusters with galaxies containing particularly immense supermassive black holes. The conventional $\Lambda CDM$ picture, on the other hand, is hard pressed to generate such extreme departures from homogeneity, owing to its requirement that the origin of inhomogeneity is taken as random quantum effects in the primordial inflaton field.
We have subsequently repeated our analysis using the more recently released Planck satellite data \cite{Planck}, and find the same general pictures of these inhomogeneities as in the WMAP data but now even more prominent than before. In Figure 1 we present the maps of temperature low-variance circle sets based on the analysis of Planck-2015 70 GHz CMB data \cite{Planck}.  The results are consistent with our previous analysis of WMAP data in \cite{Gurzadyan2013} (and with \cite{DA} though, as mentioned earlier, these authors claim compatibility with standard inflationary  $\Lambda CDM$, despite the observed very inhomogeneous sky distribution-indeed a remarkable feature of Figures 1 and 2).

\begin{figure}[!htbp]
  \centering
  \includegraphics[width=140mm]{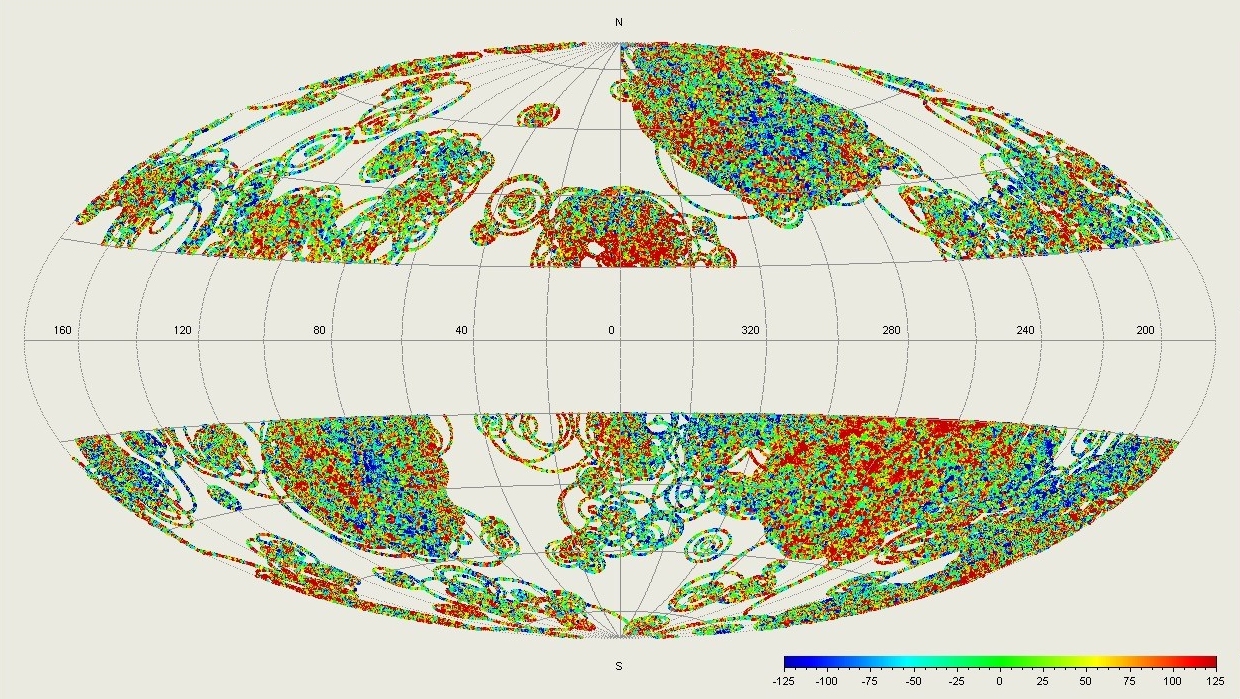}
  \includegraphics[width=140mm]{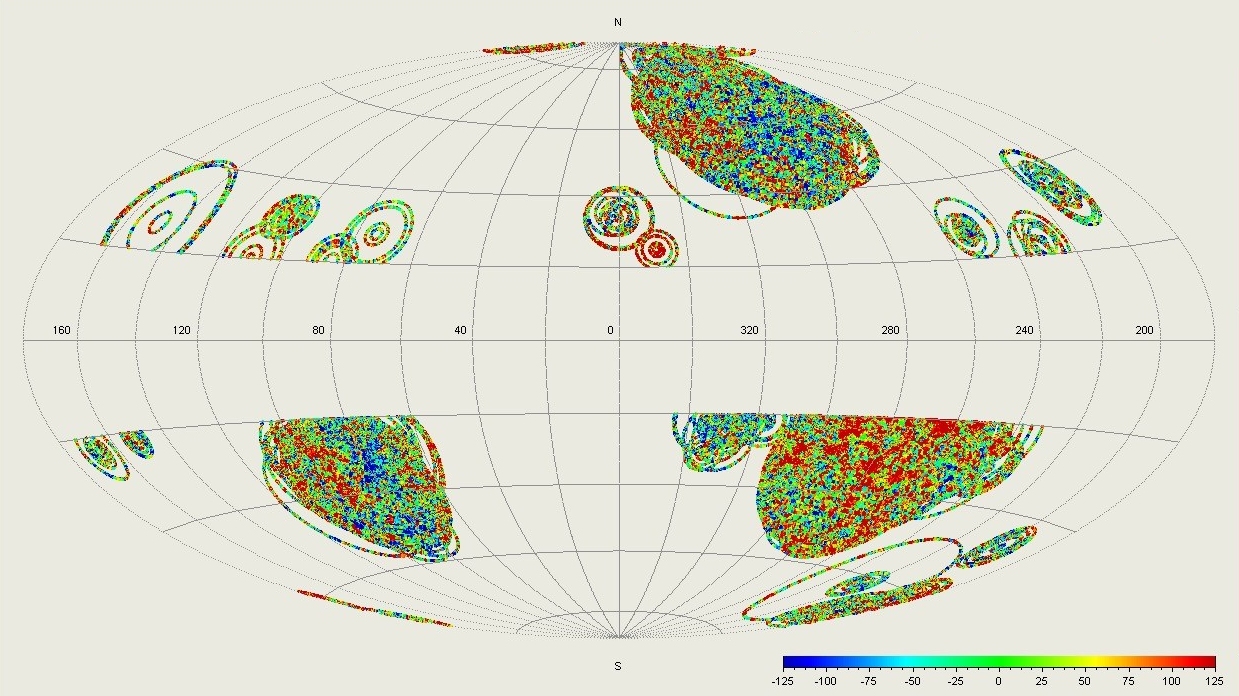}
  \label{fig:1}
  \caption{The low-variance circle sets containing 3 or more circles ({\bf a.} upper plot) and 4 or more circles ({\bf b.} bottom) of variance depth over 15 $\mu K$ (for details see \cite{Gurzadyan2013}) in Planck-2015  70 GHz data, Nside=1024.}
\end{figure}

\begin{figure}[!htbp]
  \centering
  \includegraphics[width=140mm]{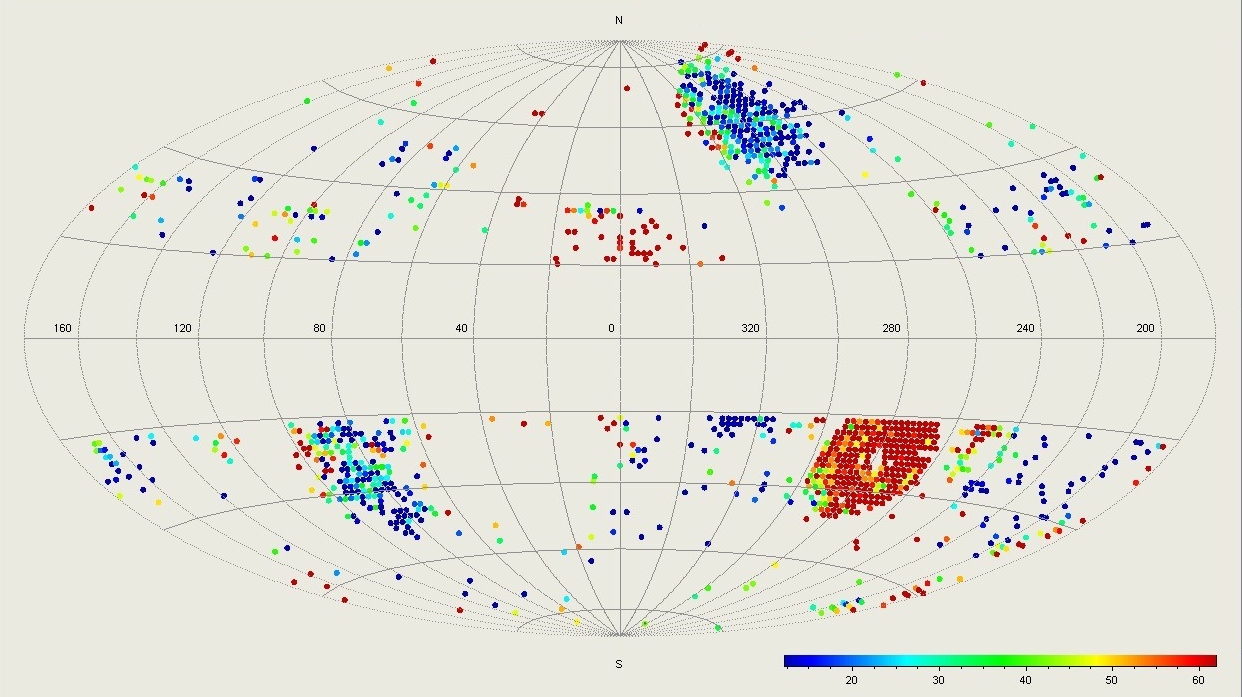}
  \label{fig:2}
  \caption{The centres of the sets in Figure 1a with assigned mean temperature of the sets.}
\end{figure}

Note that centres within the 40$^{\circ}$-wide galactic plane region are excluded. In Figure 1a, we show the locations of low-variance rings which are concentric in sets of at least 3 (finding 1134 centres), and in Figure 1b which are concentric in sets of at least 4 (finding 457 centres). This extreme inhomogeneity is particularly evident in Figure 2, where the centres themselves (for the case of Figure 1a, of  $\geq 3$-ring centres) are depicted. The colour coding shows the centres of warmer rings in red and cooler ones in blue. It is especially noteworthy, in respect of CCC, that there appears to be a strong correlation between the average temperatures of the rings and the crowding of the centres. As remarked earlier, CCC's interpretation would be that the warmer rings are blue-shifted (paradoxically shown in the conventional red in the picture) and the cooler rings (shown conventionally in blue) are red-shifted. In CCC, the more distant sources (again seemingly paradoxically) provide the blue-shifted signals, as the gravitational-wave impulse is directed towards us, so shown as red in the picture, while the relatively closer sources appear as a red-shifted signal because the gravitational-wave impulse is then directed away from us, so shown as blue in the picture. In each case we do not directly see the gravitational waves, but the equations of CCC tell us that the energy in the gravitational impulse is converted to a motion in the early dark matter distribution; hence the correlation with temperature just referred to.

\section{Where might we see signals from previous-aeon civilizations?}

In accordance with this interpretation, we conclude that, according to CCC, there was an extremely large and very distant concentration of sources, shown in red in Figure 2 just below the equatorial  excluded region, over on the right. Also, there was a comparatively close very large concentration of sources rather near the direction of the north galactic pole, just to the right of the picture. If we are to consider signals from previous-aeon beings, then such regions might well be the most promising places to look, as the CCC-interpretation would be that there might well have been vast numbers of very large galaxies in these places, and consequently a large probability of the development and long-term stability of highly evolved technological societies.

What kind of signals might we expect that such beings could be sending out? It seems highly unlikely that the manipulation of supermassive black holes would be an efficient way of sending signals, say to beyond their own aeon-even with the enormously advanced technology that might be possible for them to achieve, well before the inhospitable empty frigidity that would be the terminal situation of their aeon. From our own limited and relatively extremely primitive perspective, much more promising would undoubtedly be electromagnetic signals (although neutrinos just conceivably present us with another possibility). The conformal invariance of Maxwell's equations allow us the possibility of such signals surviving the crossover from one aeon to the next-provided that the wavelength is long enough to avoid excessive scattering by charged particles in the early stages of the subsequent aeon.

What might be a purpose to the previous-aeon beings of possibly deliberately transmitting such signals to beyond their aeon, where we must bear in mind that 2-way communication with us would be impossible in this way? Perhaps those beings might have wished to save the inhabitants of our subsequent aeon from some unpleasant fate that their greater wisdom could help us avoid. Here the purpose would, for one reason or another, simply be the transmission of information from their aeon to ours. Alternatively, there is the idea of {\it information panspermia}, introduced in \cite{Gurzadyan2005}, and attributed as "Solution 23 to Fermi paradox" in \cite{Webb}, i.e. the propagation of the "life codes" by the use of such signals, like the bit strings of human genome and of other species of terrestrial life.

In the first case, information would be transmitted with the expectation of its future decoding, perhaps for some genuinely altruistic motive. The second case can be viewed as a kind of travel by their civilization, possibly from one aeon to the next, or perhaps within a single aeon. This would be an example of what has been referred to as {\it information panspermia} \cite{Gurzadyan2005}, being based on the fact that the human genome (and that of other terrestrial species starting from bacteria, having essential common parts in their genomes) possesses low Kolmogorov complexity. (See e.g. \cite{GRS}, regarding human genome coding.) Kolmogorov complexity is defined as the minimal length of a binary coded program (in bits) required to describe the system $x$, i.e. which will enable the complete recovery of the initial system \cite{K}:
\begin{equation}
K(\phi(p),x) =min_{p:\phi=x} l(p),
\end{equation}
where $\phi(p,x)$ is a (recursive) function, calculable algorithmically by a Turing machine, and $l(p)$ is the length of the program $p$.

The corresponding bit strings might be imagined as having been transmitted, perhaps just within a single aeon, by Arecibo-type antenna over Galactic distances. One may speculate that such transmitted information, if decoded by networks of von Neumann automata or some other means, could even be equivalent to the travel of an entire civilization within an aeon, or possibly even from one CCC aeon to another. Might it be possible to eavesdrop on previous-aeon signals or even, conceivably, to reconstruct an entire previous aeon civilization? Far-fetched as such ideas may well seem, they should not be rejected out of hand, without consideration. No doubt there could well be numerous other possibilities we have not conceived of.

Finally, it is worth considering the sort of intensity of signal that previous-aeon beings might need to employ, in order to transmit something that could be observed by us. If we are considering the level of our own technology that is available to us today for making observations, and if we suppose that the beings send out their signals indiscriminately, isotropically in all directions, then we would seem to have to envisage a signal from them that harnessed such energy as the magnetic field of an entire galaxy or something of this general order. This seems hugely excessive, and we might imagine that the beings could achieve a considerable saving by aiming their signals in directions that seemed promising to them, according to their calculations as to where regions might come about in our aeon in which intelligent life might be likely to arise.  Moreover, these are very early days in our own development of a technology valuable for analyzing pre-Big-Bang signals. A far greater sophistication in this regard may be anticipated in the future, enabling us to examine the expected pre-Big-Bang signals of CCC with much greater discrimination than is available to us today - {\it whether or not} beings of the type considered above are actually sending us signals!

\section*{Acknowledgements} 
We thank the referee for a valuable comment. One of us (RP) greatly appreciates assistance from a Leverhulme Emeritus Fellowship.

\end{document}